\documentclass[english]{article}
\usepackage{mathptmx}
\usepackage[T1]{fontenc}
\usepackage[latin9]{inputenc}
\usepackage{geometry}
\usepackage{color}
\usepackage{graphicx}

\makeatletter

\usepackage{babel}
\makeatother

\begin{document}

\title{{}``Silver'' mode for heavy Higgs searches in the presence of a
fourth SM family}

\maketitle
\begin{center}
{\large S. Sultansoy}\textcolor{black}{}%
\footnote{\textcolor{black}{Institute of Physics, Academy of Sciences, Baku,
Azerbaijan.}%
}\textcolor{black}{\large , G. Unel}\textcolor{black}{}%
\footnote{\textcolor{black}{CERN, Physics Department, Geneva, Switzerland.}%
}\textcolor{black}{$^{,}$}%
\footnote{\textcolor{black}{University of California at Irvine, Physics Department,
USA. }%
} 
\par\end{center}

\begin{abstract}
We investigate the possible enhancement to the discovery of the heavy
Higgs boson through the possible fourth SM family heavy neutrino.
Using the channel $h\to\nu_{4}\bar{\nu}_{4}\to\mu\, W\,\mu\, W\to\mu\, j\, j\,\mu\, j\, j$
it is found that for certain ranges of the Higgs boson and $\nu_{4}$
masses, LHC could discover both of them simultaneously with 1fb$^{-1}$
integrated luminosity.
\end{abstract}

\section{Introduction }

The hunt for the Higgs boson is the main aim of the LHC to complete
the validation of the basic principles of the Standard Model (SM).
The main production mechanism of the Higgs boson at hadron colliders
is the gluon fusion via heavy quark loop. Therefore the number and
nature of the quarks contributing to that loop play a crucial role.
In the realm of the SM as we know it today, with 3 families, the main
contribution is from the top quark. However, the number of families
in the SM is a free parameter and the LEP-1 data fixes only the number
of fermion families with a light ($m_{\nu}<m_{Z}/2$) neutrino. The
EW precision data, on the other hand, favor the 3 and 4 family cases
equally \cite{fourthfit}. Moreover a fifth and even a sixth family
may be allowed depending on the precision measurements on $W$ boson
properties. On the other hand, the upper limit on the number of families
comes from the asymptotic freedom property of the QCD as 9.

If there is a fourth SM family, as also implied by the flavour democracy
hypothesis (see \cite{FLDem review} and references therein), its
quarks are expected to be heavier than 250 GeV, contributing to the
Higgs boson production loop in addition to the top quark. Such a contribution
enhances the production cross section for the Higgs boson and makes
the gluon fusion channel sensitive to new physics. Recently at the
Tevatron, the process $gg\rightarrow h\rightarrow WW^{*}$ is investigated
after taking into account the possible enhancement due to a fourth
SM family \cite{cdf-d0}. This mode is the most promising for the
Higgs masses between 130 and 190 GeV both at the Tevatron and the
LHC. However, its observation at the Tevatron requires the enhancement
from the fourth family.

At the LHC, in three SM family case, for a heavy Higgs of mass between
200 and 500 GeV the most prominent mode is the {}``golden'' mode
: $gg\rightarrow h\rightarrow ZZ\to\ell\ell\ell\ell$. For an even
heavier Higgs, between 500 and 800 GeV, the {}``semi-golden'' mode,
$gg\rightarrow h\rightarrow ZZ\to\ell\ell\:\nu\nu$ becomes the preferred
mode of discovery. Above 800 GeV, the discovery channel has to be
the $gg\rightarrow h\rightarrow WW\to\ell\nu jj$ \cite{R-atlas-tdr}.
The fourth SM family quarks would increase the production cross section
by a factor between 5 - 8 depending on the Higgs boson mass \cite{4th-fam-to-3rd}
decreasing the required luminosity for a 5$\sigma$ discovery.

\begin{figure}
\begin{centering}
\includegraphics[scale=0.5]{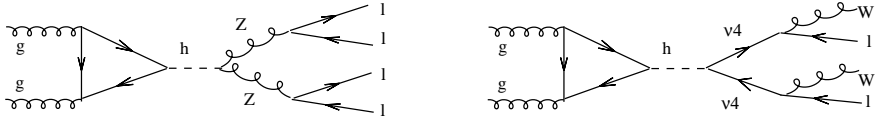}\label{fig:gold-silver}
\par\end{centering}

\caption{The {}``golden'' (left) and the {}``silver'' (right) modes for
heavy Higgs boson discovery.}

\end{figure}

The other fourth family members, depending on their mass, could also
allow new search channels for the Higgs boson. In this note, we argue
that the $gg\rightarrow h\rightarrow\nu_{4}\bar{\nu}_{4}\to\ell W\ell W$
channel, called hereafter the {}``silver'' mode, could be competitive
with the golden mode for some region of the Higgs and $\nu_{4}$ masses.
Fig. \ref{fig:gold-silver} contains the Feynman diagrams of the golden
and silver channels for the Higgs boson discovery.

\section{Fourth family neutrino pair production and decay as the {}``silver''
mode}

\begin{figure}
\begin{centering}
\includegraphics[bb=10bp 25bp 550bp 550bp,clip,scale=0.5]{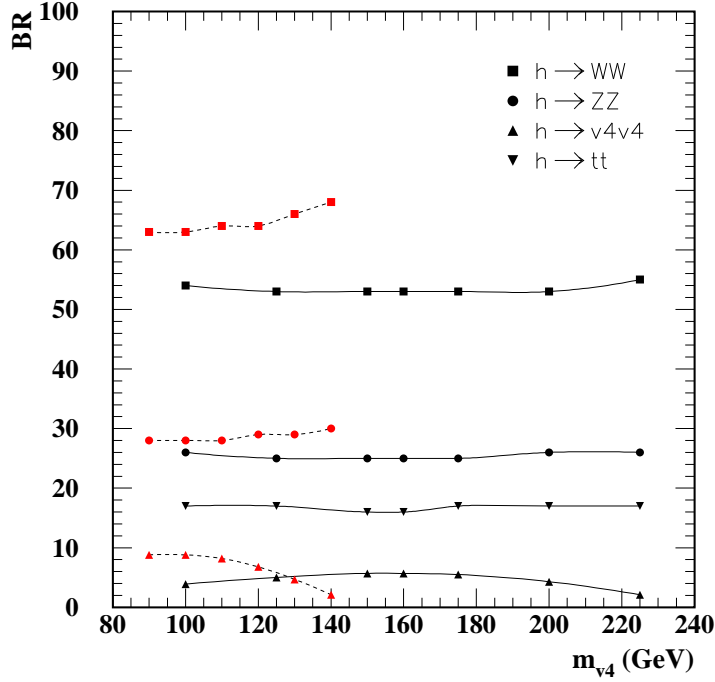}\label{fig:BR}
\par\end{centering}

\caption{The heavy Higgs branching ratios as a function of the heavy neutrino
mass for two example $m_{h}$ values 300 GeV (dashed lines) and 500
GeV (solid lines) }

\end{figure}

The fourth family neutrino, $\nu_{4}$, couples to the Higgs boson
with a vertex coefficient proportional to its mass providing a new
decay channel. The branching ratios as a function of the $\nu_{4}$
mass for two Higgs mass values, 300 and 500 GeV, are presented in
Fig. \ref{fig:BR}. As seen Br($h\rightarrow\nu_{4}\bar{\nu}_{4})$
is maximized between 90 and 100 GeV as 8.8\% for $m_{h}=$300 GeV
and between 150 and 160 GeV as 5.7\% for $m_{h}=$500 GeV. Below we
compare the {}``golden'' and {}``silver'' modes in these mass
ranges. 

The decays of the $\nu_{4}$ are governed by the leptonic 4x4 CKM
matrix. For numerical calculations, we consider the parameterization
in \cite{fourth-lepton-mixing} which is compatible with the experimental
data on the masses and the mixings in the SM leptonic sector. In this
case, the Br($\nu_{4}\rightarrow\mu W)\simeq0.68$ and Br($\nu_{4}\rightarrow\tau W)\simeq0.32$
which imposes the main discovery signal as two muons and four jets
considering the hadronic decays of the $W$ bosons in the final state.
Note that the {}``silver'' mode contains only muons compared to
both electrons and muons of the {}``golden'' mode. This scenario
with $m_{h}=300$ GeV, leads to Br($h\rightarrow\nu_{4}\bar{\nu}_{4}\rightarrow\mu^{+}\mu^{-}\, jjjj)$=1.22$\times10^{-2}$
which should be compared to the {}``golden'' mode branching ratio
of 1.12$\times10^{-3}$, giving an enhancement factor about 11. Corresponding
numbers for $m_{h}=500$ GeV are 1.88$\times10^{-2}$ and 1.25$\times10^{-3}$
respectively, yielding an enhancement of about 15 times. We believe
that an order of magnitude higher statistics would compensate the
possible inefficiencies associated with jet detection and hadronic
$W$ reconstruction. 

An associated channel to the {}``silver'' mode is the case where
one of the $W$ boson decays leptonically: $W\to\ell\,\nu$ ($\ell=\mu,e$).
The final state in this case will be $\mu^{+}\mu^{-}\,\ell jjE\!\!\!\!\!/_{T}$
. The number of such events is 63\% of the {}``silver'' mode discussed
above, bringing the total enhancement factor up to 24 (18) compared
to the {}``golden'' mode for a Higgs boson of $m_{h}$=500 (300)
GeV. 

If the fourth family neutrino is of Majorana nature, an experimentally
clear signature would be available, namely same sign muons as decay
products of $\nu_{4}$s. Although in this case, the number of expected
signal events is halved, the SM background is practically negligible
making this mode deserve the name {}``platinum'' mode.

\section{Conclusion}

If Nature allows, a double discovery in the first year of the LHC
start up is in the realm of the possible: the fourth family neutrino
and a heavy ($m_{h}>300$ GeV) Higgs boson. For $m_{h}=300$ (500)
GeV the fourth family quarks increase the Higgs production cross section
to 7$\times$10$^{4}$ ( 2.5$\times10^{4}$) fb  compared to $10^{4}$
($5\times10^{3}$) fb in the 3 family SM case \cite{higgs-prod}.
Consequently, the so called {}``silver'' mode allows about 850 (470)
Higgs bosons (and obviously twice as many $\nu_{4}$) to be reconstructed
with 1fb$^{-1}$ luminosity for $m_{h}=$300 (500) GeV and $m_{\nu_{4}}=$100
(150) GeV. The Monte Carlo simulation to verify this statement is
under progress.

\subsection*{Acknowledgments}

S. S. would like to thank P. Jenni for the provided support during
his visit to CERN. G. U.'s work is supported in part by U.S. Department
of Energy Grant DE FG0291ER40679. S. S. is also grateful to Gazi University
science and letters faculty deanship for relieving him from his teaching
duties. The authors would like to thank M. Karagoz-Unel for useful
comments.

\textbf{Note added:} After the submission of the first version of
this note, an arXiv entry made one month ago, mentioning an enhancement
to Higgs discovery from $h\to\nu_{4}\bar{\nu}_{4}\to\ell\, W\,\ell\, W\to4\ell+E\!\!\!\!\!/_{T}$
channel was brought to our attention \cite{tait}.

\end{document}